\documentstyle[prl,aps,epsf,here]{revtex}

\renewcommand{\arraystretch}{1.05}
\def\be{\begin{equation}}
\def\ee{\end{equation}}
\def\bea{\begin{eqnarray}}
\def\eea{\end{eqnarray}}
\def\eq#1{(\ref{#1})}
\def\tab#1{Table \ref{#1}}
\def\bs{\bigskip}
\def\ms{\medskip}
\def\fig#1{Fig.\ \ref{#1}}
\def\eq#1{(\ref{#1})}

\def\siml{\,\hbox{\kern.1em \lower.6ex \hbox{$\sim$} \kern-1.12em
          \raise.6ex \hbox{$<$} }}
\def\simg{\,\hbox{\kern.1em \lower.6ex \hbox{$\sim$} \kern-1.12em
          \raise.6ex \hbox{$>$} }}
\def\d{{\rm d}}
\newcommand{\Figurebb}[9]{
\begin{figure}[H]\begin{center}
\leavevmode
\epsfysize=#7cm
\epsfbox[#2 #3 #4 #5]{#6}
\par
\parbox{#8cm}{
\caption[figure]{\renewcommand{\baselinestretch}{0.8} \small
                                           \hspace{-0.3truecm}#9}
\label{#1}}\end{center}
\end{figure}
}

\begin{document}
\draft

\title{Periodic orbit theory for the H\'enon-Heiles system 
       in the continuum region}
\author{J. Kaidel, P. Winkler$^1$ and M. Brack}
\address{Institut f\"ur Theoretische Physik, Universit\"at Regensburg,
         D-93040 Regensburg, Germany\\
         $^1$Department of Physics, University of Nevada
         Reno, NV 89557, USA}
\date{\today}
\maketitle

\vspace*{0.6cm}

\begin{abstract}
We investigate the resonance spectrum of the H\'enon-Heiles potential up to 
twice the barrier energy. The quantum spectrum is obtained by the method of 
complex coordinate rotation. We use periodic orbit theory to approximate
the oscillating part of the resonance spectrum semiclassically and Strutinsky
smoothing to obtain its smooth part. Although the system in this energy 
range is almost chaotic, it still contains stable periodic orbits. 
Using Gutzwiller's trace formula, complemen\-ted by a uniform approximation for 
a codimension-two bifurcation scenario, we are able to reproduce the 
coarse-grained quantum-mechanical density of states very accurately, including 
only a few stable and unstable orbits.
\end{abstract}


\section{Introduction}

Understanding the way in which scattering resonances modify the density of 
states in the continuum region of a quantum-mechanical system has been a 
challenging problem since the early days of quantum mechanics \cite{bet37}. 
Processes involving quantum resonances are ubiquitous in nature and 
technology \cite{fey}. They contribute, e.g., to the conductance 
fluctuations in transport phenomena \cite{bak}. The focus of our present
paper lies on the semiclassical evaluation of the density of states
of an open system. The periodic orbit theory developed over 35 years 
ago by Gutzwiller \cite{gutz}, and its extension to closed orbits 
\cite{delos}, have had an enormous impact on the recent research on 
``quantum chaos'', i.e., the study of quantum signatures of classical chaos 
\cite{gbook,stock,haake}. Numerous studies have shown that resonance spectra 
can also be approximated semiclassically, using either the 
closed or the periodic orbits of the underlying classical system. Besides 
efforts which were limited to fully chaotic systems \cite{gas89,wir}, recent 
interest has focused on general systems with mixed phase-space dynamics, 
including potentials which do not vanish asymptotically \cite{tan96,kon,bar}. 
However, a satisfactory semiclassical description of resonance 
spectra could only be achieved in those limits where all orbits are unstable 
and can be completely enumerated by symbolic dynamics. Truly mixed-dynamical
systems, in which unstable and stable orbits coexist and undergo bifurcations,
pose serious problems for the semiclassical theories. Although the divergences
arising at bifurcations can be remedied by uniform approximations
for the simplest codimension-one \cite{ozoha,ssuni} and codimension-two 
scenarios \cite{schom,scho2,jkmb}, their enormous proliferation with increasing 
orbit length renders a semiclassical determination of the fine structure of 
quantum spectra practically impossible.

Presently we study the two-dimensional H\'enon-Heiles (HH) potential \cite{hh64}
which is a well-known paradigm for a mixed system \cite{wal69} and has served 
as a model for various physical systems of different nature 
\cite{ll,bog96,pom74,efs04}. This paper is the sequel of a recent semiclassical 
study of the HH potential up to the barrier energy \cite{jkmb}. Here we shall 
determine the quantum-mechanical resonance spectrum above the barrier, extract 
the smooth density of states by Strutinsky averaging \cite{stru}, and calculate 
its oscillating part via Gutzwiller's trace formula \cite{gutz}, incorporating a 
uniform approximation to regularize a codimension-two sequence of periodic orbit 
bifurcations. 

While the fine structure of the spectrum is not accessible semiclassically for 
the reasons stated above, we consider here the {\it coarse-grained} density of 
states obtained by a Gaussian convolution over a finite energy range. This 
allows one to include only a finite number of shorter orbits but still to 
reproduce the gross-shell structure of the level density \cite{strma}, as has 
been exemplified in various models and physical applications (see, e.g., 
\cite{bbuch}). For the HH potential, this was shown in the low-energy 
range \cite{hhuni} using a uniform treatment of the SU(2) symmetry limit, and 
for energies close to the barrier \cite{jkmb} using a uniform treatment of 
sequences of pitchfork bifurcations. Here we show that also in the continuum 
region above the barrier, the coarse-grained quantum-mechanical density of 
states is very well reproduced semiclassically using a relatively small number 
of unstable and stable periodic orbits.

\section{Density of states including the continuum}

Let us consider a particle scattered by a spherical\-ly symmetric one-body 
potential $V(r) < 0$ with $V(r)\rightarrow 0$ for $r\rightarrow\infty$. 
The density of states of the free system in the continuum region, given by
\be 
g_{free}(E) = c\,\sqrt{E} \qquad (E>0)
\label{gfree}
\ee 
with constant $c$, is modified through the scattering resonances by a 
contribution \cite{bet37}
\begin{equation}
\Delta g\left(E\right)=\frac{1}{\pi}
                             \sum_{l=0}^\infty \frac{\partial\,\delta_{l}(E)}
                             {\partial E}\,. \qquad (E>0)
\label{Delg}
\end{equation}
Here $l$ are the quantum numbers of the orbital angular momentum and 
$\delta_{l}(E)$ is the elastic scattering phase shift of the $l$-th 
partial wave. By definition, resonances occur at those energies $E_{l}$ 
where the phase shift takes the value $\delta_{l}(E_{l})=\pi/2$. Expanding 
the phase shift around the res\-onance energy one obtains
\begin{equation}
\delta_{l}(E) = \arctan\left(\frac{\Gamma_{l}/2}{E-E_{l}}\right) + \;\dots,
\label{phase}
\end{equation}
where $\Gamma_{l}$ is the width of the resonance, related to its life 
time $\tau_{l}$ by $\Gamma_{l}=\hbar/\tau_{l}$. Inserting \eq{phase} 
into \eq{Delg}, keeping only the leading term, leads to
\begin{equation}
\Delta g\left(E\right)=\frac{1}{\pi}\sum_{l=0}^\infty 
                       \frac{\Gamma_{l}/2}{\left(E-E_{l}\right)^2
                       +\left(\Gamma_{l}/2\right)^2} \,.
\label{DelgG}
\end{equation}
In the region $E<0$, where the potential has only discrete eigenvalues
$E_{nl}$ with a radial quantum number $n$, the density of
states is given by a sum of delta functions and the total
density of states for the system is given by
\be
g_{tot}(E) = g_{free}(E) + \!\sum_{n,l=0}^\infty 
             \delta(E-E_{nl}) + \Delta g(E)\,. 
\label{gtot}
\ee
For a non-integrable system without spherical symmetry, the spectrum 
of both bound states and resonances can only be characterized by one
quantum number, say $m$, replacing $(n,l)$ in the above. We thus
rewrite \eq{DelgG} as
\bea
\Delta g(E) = g_{tot}(E)-g_{free}(E) 
            = \frac{1}{\pi}\sum_m\frac{\Gamma_{\!m}/2}{\left(E-E_m\right)^2
                  +\left(\Gamma_{\!m}/2\right)^2}\,,
\label{dos}
\eea 
whereby the bound spectrum is automatically included since the Lorentzians 
on the r.h.s.\ go over into delta functions for $\Gamma_{\!m}\rightarrow 0$.

We next define a {\it coarse-grained} density of states, performing a 
Gaussian convolution of \eq{dos} over an energy range $\gamma$. This can be 
done analytically, leading to
\begin{eqnarray}
\Delta g_{\gamma}(E) 
      = \frac{1}{\gamma\sqrt{\pi}}\int_{-\infty}^\infty
            \Delta g(E')\, e^{-(E-E')^2/\gamma^2}\, \d E' 
      = \frac{1}{\gamma \sqrt{\pi}} \sum_m
            \Re e \left[w(z_m)\right],
\label{ggam}
\end{eqnarray}
with
\be
w(z) = e^{-z^2} {\rm erfc}(-iz)\,,\quad
z_m = \frac{(E_m+i\,\Gamma_{\!m}/2-E)}{\gamma}, 
\label{erf}
\ee
where ${\rm erf}(z)=1-{\rm erfc}(z)$ is the error function \cite{abr}.

\newpage

\section{Numerical calculation of the H\'enon-Heiles spectrum}

For a general potential $V({\bf r})$ without spherical symmetry, it may
become difficult to calculate the scattering phase shifts. Furthermore,
if the potential has a continuous spectrum above some threshold energy 
$E_{th}$ but does not reach $E_{th}$ asymptotically for 
$r\rightarrow\infty$ (such as the H\'enon-Heiles potential considered below), 
there are generally no free plane-wave solutions 
for $E>E_{th}$ and the phase shifts cannot be defined. Nevertheless, there
are ways to calculate complex resonance energies $E^*_m=E_m-i\,\Gamma_{\!m}/2$
which appear as poles of the Green function in the complex energy plane
$E^*$ with $\Re e E^*>E_{th}$.

One convenient way to obtain the resonances without requiring the knowledge
of phase shifts is given by the method of complex rotation 
\cite{bal71,sim73,yar78,moi98}. Here one solves the scaled Schr\"odinger 
equation
\begin{equation}
\left[\hat{S} \hat{H}({\bf r}) \hat{S}^{-1}\right] 
\hat{S} \phi^{res}_m \left({\bf r}\right) =
\left(E_m-i\,\Gamma_{\!m}/2\right)
\hat{S} \phi^{res}_m \left({\bf r}\right),
\label{rotSG}
\end{equation}
where $\hat{S}$ is the similarity transformation (or complex rotation)
\begin{equation}
\hat{S} f\!\left({\bf r}\right)=f\!\left({\bf r} e^{i \theta}\right)
\end{equation}
which multiplies each spatial coordinate of an analytical function 
$f\!\left({\bf r}\right)$ by a complex exponential with real phase
$\theta$. This transformation turns a resonance wave\-function 
$\phi^{res}_m\left({\bf r}\right)$ into a square-integrable function which 
can be expanded in Hilbert space \cite{moi98}. For systems with 
asymptotically free states, the energies $E>E_{th}$ of all nonresonant 
continuum states are rotated in the complex plane to the line 
$(E-E_{th})\exp(-2i\theta)$, whereas the poles at $E^*_m=E_m-i\,\Gamma_{\!m}/2$ 
corresponding to the resonant states remain independent of $\theta$, 
provided that this angle is large enough to ``uncover'' the poles, i.e., 
$2\theta>\arctan\,[\Gamma_{\!m}/2(E_m-E_{th})]$. Practical experience shows 
\cite{alpot} that $\theta$-independent poles can also be found if the 
non-resonant continuum states are not asymptotically free, although this 
has not been proven rigorously. Note that the discrete eigenenergies in the 
bound region $E<E_{th}$ are also obtained by the complex rotation method; 
they stay on the real energy axis with $E<E_{th}$ and have zero imaginary 
parts.

Having determined the energies $E^*_m=E_m-i\,\Gamma_{\!m}/2$, including the
bound spectrum with $\Gamma_{\!m}=0$, their contribution to the density of 
states is given by
\be
\Delta g(E) = -\frac{1}{\pi}\,\Im m\sum_m\frac{1}{E-E_m+i\,\Gamma_{\!m}/2},
\ee 
leading to \eq{dos}.

We now want to investigate the density of states of the two-dimensional 
H\'enon-Heiles (HH) Hamiltonian 
\begin{equation}
\hat{H}=-\frac12\!\left({\hat p}_x^2+{\hat p}_y^2\right)
         +\frac12\!\left(x^2+y^2\right)
         +\alpha\,(x^2 y-y^3\!/3)\,, 
\label{rH}
\end{equation}
with $\alpha>0$ and units such that $\hbar=1$. This Hamiltonian describes 
an open system in which a particle can escape by direct transmission over 
-- or by tunneling through -- one of three barriers with height 
$E_{bar}=1/6\alpha^2$. However, since the potential goes asymptotically to 
$-\infty$ like $-r^3$ (with $r^2=x^2+y^2$) in some regions of space, the 
system has no discrete eigenstates and we must put $E_{th}=-\infty$. For 
sufficiently small values of the parameter $\alpha$, there are quasi-bound 
states below the three barriers, but they have finite widths due to tunneling.

Scaling both coordinates and momenta with the factor $1/\alpha$ causes the 
classical dynamics to depend only on the scaled energy 
$e=E/E_{bar}=6\alpha^2 E$; the barrier energy then lies at $e_{bar}=1$. 
In the following we present all real parts of the spectrum in the scaled 
energy units $e$.

To solve the complex Schr\"odinger equation \eq{rotSG} for the system \eq{rH}, 
we diagonalize it in a truncated basis $\left\{| n m \rangle\right\}$ of the 
two-dimensional isotropic harmonic oscillator with $n,m\leq N$. This leads to 
the eigenvalue problem for the complex non-Hermitian matrix 
\begin{eqnarray}
\left[H\left(\theta\right)\right]_{n'm'nm} &\equiv&
\langle n'm'\left|\hat{S} \hat{H}\left({\bf r}\right) \hat{S}^{-1} 
\right| nm\rangle \nonumber \\
&=& T_{n'm'nm}\, e^{-2i \theta}+\left(V_{HO}\right)_{n'm'nm} e^{2i \theta}
 +\,\alpha \left(V_3\right)_{n'm'nm} e^{3i \theta}.
\label{cH}
\end{eqnarray}
The matrices $T_{n'm'nm}$, $\left(V_{HO}\right)_{n'm'nm}$ and 
$\left(V_3\right)_{n'm'nm}$ are the real matrices of the kinetic, harmonic 
and cubic parts of \eq{rH}, respectively, for $\theta=0$. The complex
eigenvalues $E^*_m=E_m-i\Gamma_m$ of the matrix \eq{cH} were found
numerically using its sparse property. 
 
\Figurebb{HHcspec}{120}{495}{378}{693}{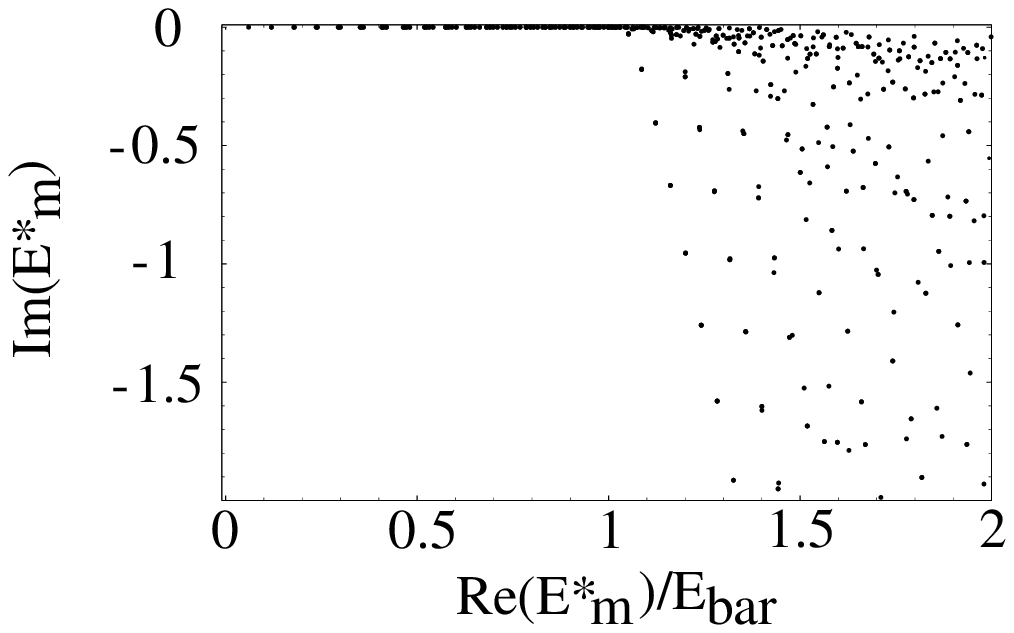}{3.9}{14.5}{
Resonances $E^*_m=E_m-i\Gamma_m$ of the H\'enon-Heiles system with 
$\alpha=0.1$: imaginary parts versus real parts (the latter in scaled
energy units $e=E/\!E_{bar}$). 
}

In \fig{HHcspec} the complex resonance spectrum of the HH system
\eq{rH} for $\alpha=0.1$ is shown. The size of the truncated basis 
$| n m \rangle$ was given by $N=n_{max}=m_{max}=130$. We only give 
the spectrum above the minimum of the classically bound region, located at 
$e=0$. Due to the truncation of the basis in Hilbert space, it is known 
\cite{yar78} that the poles in the complex energy plane slightly depend on 
the rotation angle $\theta$; their optimal values are then found as 
stationary points (or plateau values) with respect to small variations of
$\theta$. We could determine the plateau values of the resonances with an 
accuracy of 6 digits over an interval in $\theta$ of about 10 degrees, as 
is shown by the example in \fig{specplat}. (The real part of the
resonance, $E_m/\!E_{bar}=0.928966$, is constant within 6 digits in the 
whole interval of $\theta$ shown.) Note that the imaginary parts of the 
quasi-bound states for $e<1$ are exponentially small except very near the 
barriers. For the states slightly above the barriers, a semiclassical 
prediction of the imaginary parts, which is in good agreement with our 
numerical results, was given in \cite{mil}.
The quasi-regular pattern observed in the region $e>1$, where some
of the resonances lie on almost parallel ``rays'' in the complex
energy plane, is a reminiscence of the separable system that is obtained
if one neglects the coupling term $\alpha x^2y$ in \eq{rH} (see
Ref.\ \cite{jkmb} for the density of states of this separable system). 

\Figurebb{specplat}{130}{178}{425}{380}{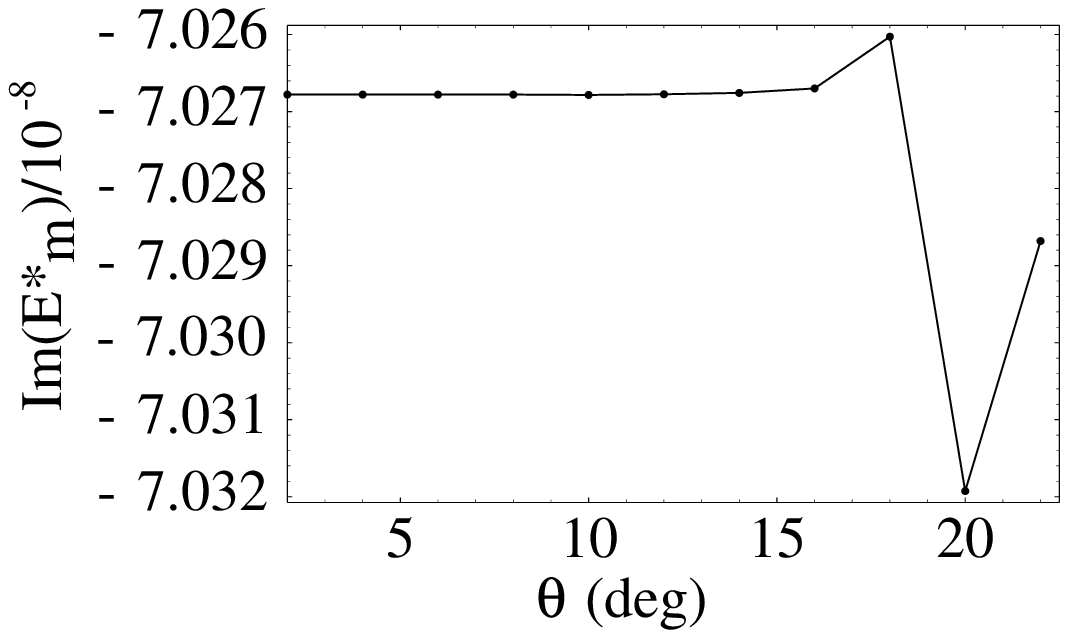}{3.7}{14.5}{
Imaginary part $-\Gamma_m$ of a resonance with real part $E_m/\!E_{bar}=
0.928966$, plotted versus the rotation angle $\theta$.
}

\section{Smooth part of density of states}
\label{secstru}

The main purpose of our paper is to establish the classical-to-quantum
correspondence, approximating the density of states \eq{dos} by a
semiclassical trace formula \cite{gutz}. As usual, the trace formula only 
yields the oscillating part $\delta g_\gamma(E)$ of the total density of 
states, written as
\be
g_{tot}(E) = g_{free} + {\widetilde g}(E) + \delta g_{\gamma}(E)\,,
\ee
where $g_{free}$ in the present two-dimensional system is a constant.
The function ${\widetilde g}(E)$ is the smoothly varying non-periodic part 
of \eq{dos}, which usually is obtained from the (extended) Thomas-Fermi 
(ETF) model \cite{bbuch}. In the present HH system, however, we have the 
problem that the ETF level density cannot be defined for $e>1$ where the 
system is open. We therefore resort to the numerical Strutinsky averaging 
\cite{stru} which is equivalent to the ETF model where the latter can 
be used \cite{bbuch}. As shown in \cite{bra72}, the Strutinsky averaging 
corresponds to approximating the average part of the density of states near 
the energy $E$ by a polynomial of given power $2s$, i.e., a truncated Taylor 
expansion with $s=1,2,\dots$ In practice it is obtained by convolution of 
the density of states with a Gaussian of width $\,\tilde{\!\gamma}$, 
modified by a suitable linear combination of Hermite polynomials up to order 
$2s$. The smoo\-thing function can also be compactly written as \cite{ros72}
\be
f_{\,\tilde{\!\gamma},s}(E) = \frac{1}{\,\tilde{\!\gamma}\sqrt{\pi}}\,
                              e^{-(E/\,\tilde{\!\gamma})^2}
                              L^{1\!/2}_s
                              \left[(E/\,\tilde{\!\gamma})^2\right],
\label{fgam}
\ee 
where $L^{1/2}_s$ is an associated Laguerre polynomial. The
average part of \eq{dos} is then obtained by the convolution
\be
{\widetilde g}(E) = \int_{-\infty}^\infty \Delta g(E')\,
                    f_{\,\tilde{\!\gamma},s}(E-E')\, \d E'\,.
\label{gav}
\ee
Ideally, the results obtained in this way will not depend on $s$ and 
$\,\tilde{\!\gamma}$, provided that $\,\tilde{\!\gamma}$ is chosen to be 
larger than the characteristic energy spacing of the main shells in
the spectrum and $s$ is large enough. This is indeed the case if the
true average (ETF) density of states is a polynomial of order $\leq 2s$. 
Practically, one has to look for stationary values of the results as
functions of both $\,\tilde{\!\gamma}$ and $s$, fulfilling the so-called
``plateau condition'' \cite{bra72}. The integral in \eq{gav}, with 
$\Delta g(E)$ given by \eq{dos}, can be calculated analytically (see
Appendix).

\Figurebb{plhh}{110}{450}{392}{646}{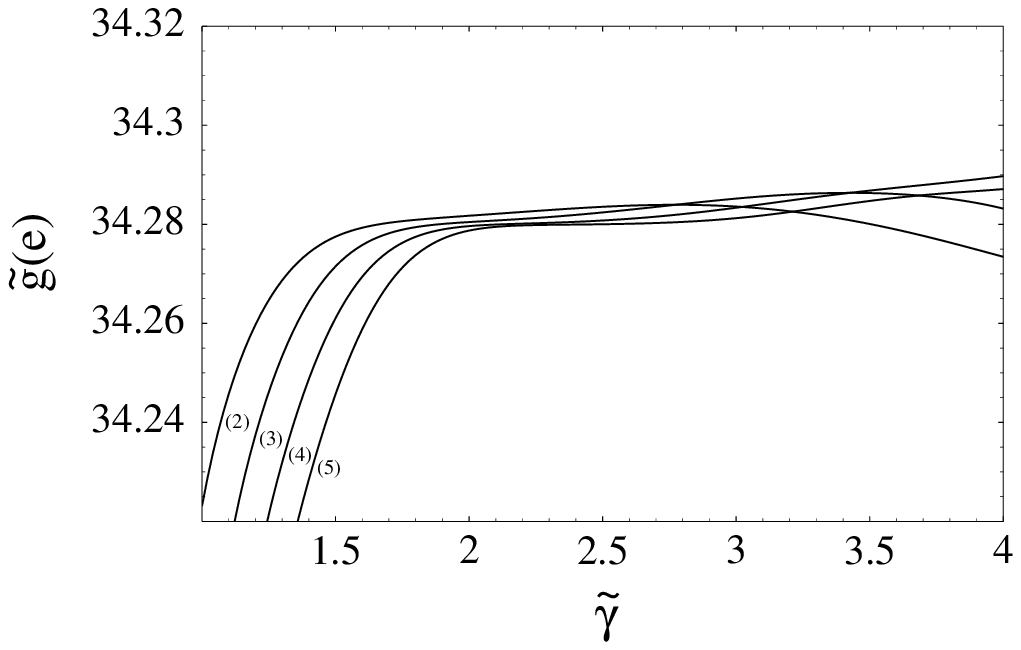}{4.8}{14.5}{
Test of the plateau condition for the average density of states 
of the HH system with $\alpha=0.1$, evaluated at the 
energy $e=1.5$. The numbers in brackets give the order $s$ of the 
Laguerre polynomial in \eq{fgam}.}

In \fig{plhh} we show the plateaux in ${\widetilde g}(E)$ obtained by 
varying $\,\tilde{\!\gamma}$ and the polynomial order $s$ at the fixed energy 
$e=1.5$. One can see that the stationary condition is reasonably well 
fulfilled for $\,\tilde{\!\gamma} \approx 2.2$, independently of $s>2$.
Plateaux of this quality have been obtained for the spectra of finite-depth
Woods-Saxon potentials appropriate for nuclear physics, and the 
plateau values of the averaged quantitities have been shown to be identical 
with their ETF values within the numerical accuracies \cite{sob}.

Having determined the optimal plateau values at all energies of interest,
the quantum-mechanical value of $\delta g_\gamma(E)$ is given by
\be
\delta g_\gamma(E) = g_\gamma(E) - {\widetilde g}(E)\,.
\ee
In the HH system, the function $\widetilde{g}\left(E\right)$ 
varies rather abruptly near $e=1$ on a scale comparable to the oscillations 
of $\delta g_\gamma\left(E\right)$, so that no ideal plateaux are found and 
there remains a small numerical uncertainty near $e\simg 1$. 

\section{Semiclassical calculation of the coarse-grained resonance spectrum}

Next we want to construct the semiclassical approximation of 
$\delta g_\gamma\left(E\right)$ in the form of Gutzwiller's trace formula
for isolated orbits \cite{gutz},
modified by the exponential factor which is the result of the 
coarse-graining over the energy range $\gamma$ and suppresses the 
contributions from orbits with longer periods: 
\bea
\delta g_{scl}\left(E\right) = 
\frac{1}{\pi \hbar} \sum_{\xi} \frac{T_{\xi}(E)}
{r_{\xi}\sqrt{\left|{\rm Tr} \widetilde{M}_{\xi}(E)-2\right|}}\,
e^{-[\gamma T_{\xi}(E)/2\hbar]^2} 
    \cos\left[\frac{S_{\xi}(E)}{\hbar}-\frac{\pi}{2}\sigma_{\xi}\right],
\label{dggamma}
\eea
The sum goes over all isolated
periodic orbits labeled by $\xi$, and the other quantities in \eq{dggamma}
are, as usual, the periods $T_\xi$ and actions $S_\xi$, 
the Maslov indices $\sigma_\xi$ and the repetition numbers $r_{\xi}$ 
of the periodic orbits. $\widetilde{M}_{\xi}(E)$ are the stability matrices
obtained by linearization of the equations of motion along the periodic 
orbits.

The shortest periodic orbits of the classical HH system \eq{rH} were 
obtained using a numerical Newton-Raphson algorithm \cite{kaid}. They 
have already been extensively studied in earlier papers 
\cite{chu,dav,vioz,mbgu,lame}. We use here the nomenclature introduced in 
\cite{lame}, where the Maslov indices $\sigma_{\xi}$ appear as subscripts in 
the symbols (B$_4$, R$_5$, L$_6$, etc.) of the orbits. The Maslov indices 
were obtained by the method developed in \cite{cre} (see Ref.\ \cite{bbuch} 
for practical recipes). 

\newpage

\Figurebb{hhorbs}{200}{425}{392}{685}{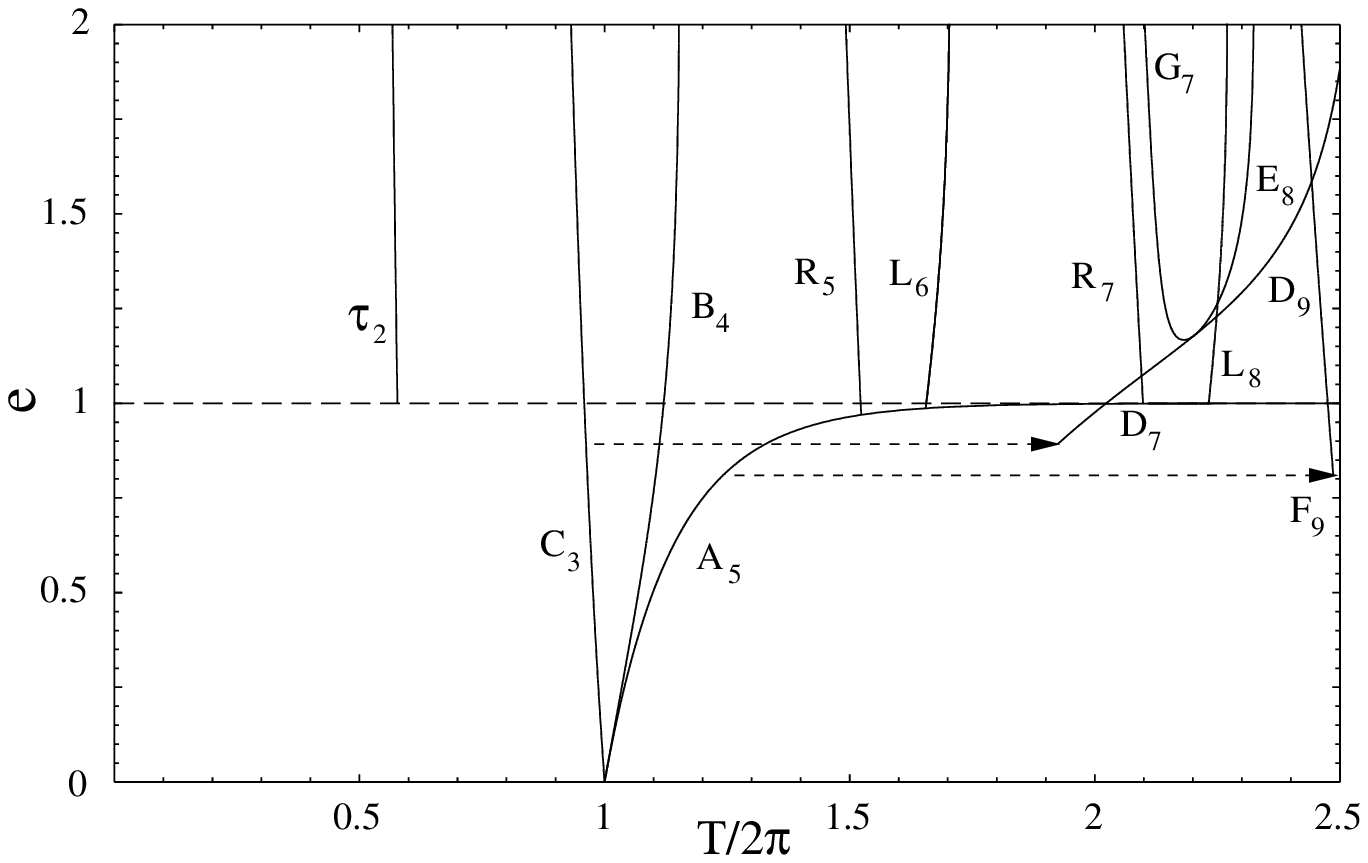}{6.8}{14.5}{
Scaled energy $e$ versus periods $T_\xi/2\pi$ of the shortest periodic orbits 
in the HH potential. The dotted arrows correspond to period-doubling 
bifurcations.}

\vspace*{-0.1cm}
\fig{hhorbs} shows the periods of the shortest periodic orbits as functions 
of the scaled energy $e$ of the system up to twice the barrier energy. One 
can see that there still exist many periodic orbits above the barrier 
energy where the particle has enough energy to escape from the bound region. 
Actually, there is an infinite number of orbits of type R and L (only the 
two shortest of each are shown here), born from the saddle-line orbit A in 
a cascade of bifurcations \cite{dav,mbgu,lame} cumulating at $e~\!=~\!1$. They 
exist at all energies $e$ above their respective bifur\-cations but become 
very unstable at higher energies. Above the barriers ($e>1$), new orbits 
(named S in \cite{dav} and $\tau$ in \cite{vioz,mbgu}) with Maslov index 2 
arise, librating across the saddles. Although these $\tau_2$ orbits are quite 
unstable, they have the smallest periods of all orbits and therefore play an 
important role for the coarse-grained density of states at $e>1$, as discussed 
in the following.

Here we concentrate on the density of states above the barriers ($e\geq 1$). 
For semiclassical calculations at $e<1$, we refer to earlier papers 
\cite{jkmb,hhuni}. The periodic orbits in the region $e>1$ are not all 
unstable. This can be inferred from \fig{hhorbs} for the bifurcation of the 
stable orbit D$_7$/D$_9$ near $e\simeq 1.18$, in which the orbit E$_8$ and, 
indirectly, also the orbit G$_7$ is involved. It is seen directly in 
\fig{trmbif}, where we show the traces of their stability matrices versus $e$. 

\vspace*{-0.1cm}
\Figurebb{trmbif}{67}{287}{502}{540}{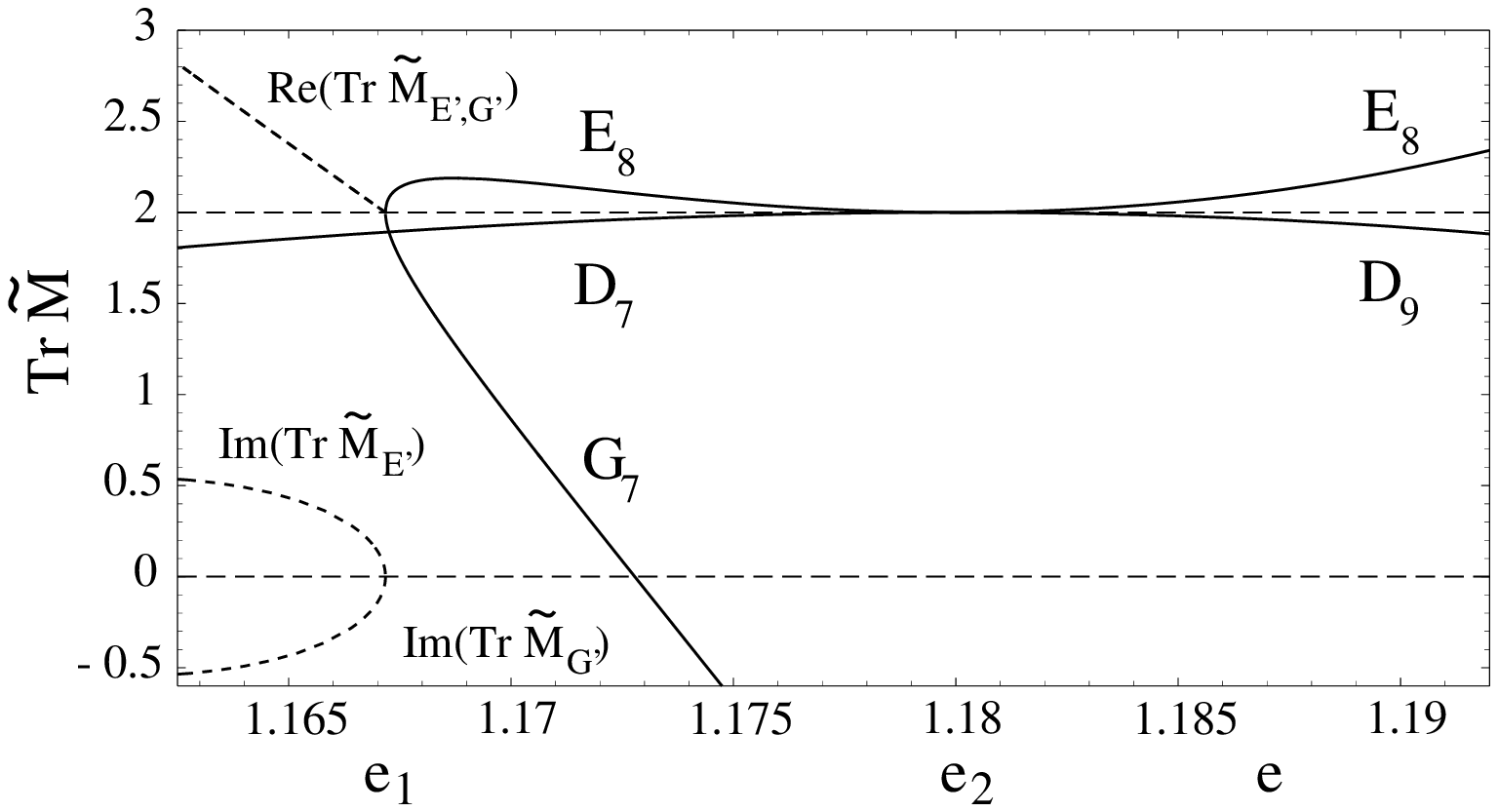}{4.7}{14.5}{
Trace of stability matrix $\widetilde M_\xi$ vs.\ scaled energy $e$ for 
the period-two orbits $\xi$ = D$_7$/D$_9$, E$_8$, and G$_7$ taking part 
in a codimension-two sequence of bifurcations. The short-dashed lines show 
the values for the ghost orbits E' and G' associated to the tangent 
bifurcation. (Upper left: common real part, lower left: imaginary parts of 
tr$\widetilde M_{E'}$ and tr$\widetilde M_{G'}$.) 
}
\vspace*{-0.1cm}
In \fig{xybif} we show the shapes of these three orbits at three energies
around $e\simeq 1.18$  where the D and E orbits meet. Note that the central 
D orbit (solid lines) has $C_3$ symmetry. The satellite orbits E$_8$ 
(dashed lines) and G$_7$ (dash-dotted lines) do not have this symmetry and
occur in three orientations (and hence must be included thrice in the
trace formula). All three orbits have an additional degeneracy of two from 
time reversal symmetry.

This scenario represents a sequence of bifurcations corresponding to an 
unfolding of codimension two: at $e_1=1.16717$, the orbits E$_8$ and 
G$_7$ are born in a tangent bifurcation; then the unstable orbit E$_8$
and the stable orbit D$_7$ meet at $e_2=1.18000$ in a touch-and-go
bifurcation, leaving it as D$_9$ and E$_8$. The two bifurcations are 
too close to be treated separately as codimension-one scenarios, because
the action differences of the orbits involved between the two bifurcation 
energies, $\Delta S_\xi = S_\xi(e_2)-S_\xi(e_1)$, do {\it not} fulfill the 
requirement $|\Delta S_\xi|\gg \hbar$.

\vspace*{-0.55cm}
\Figurebb{xybif}{0}{0}{724}{254}{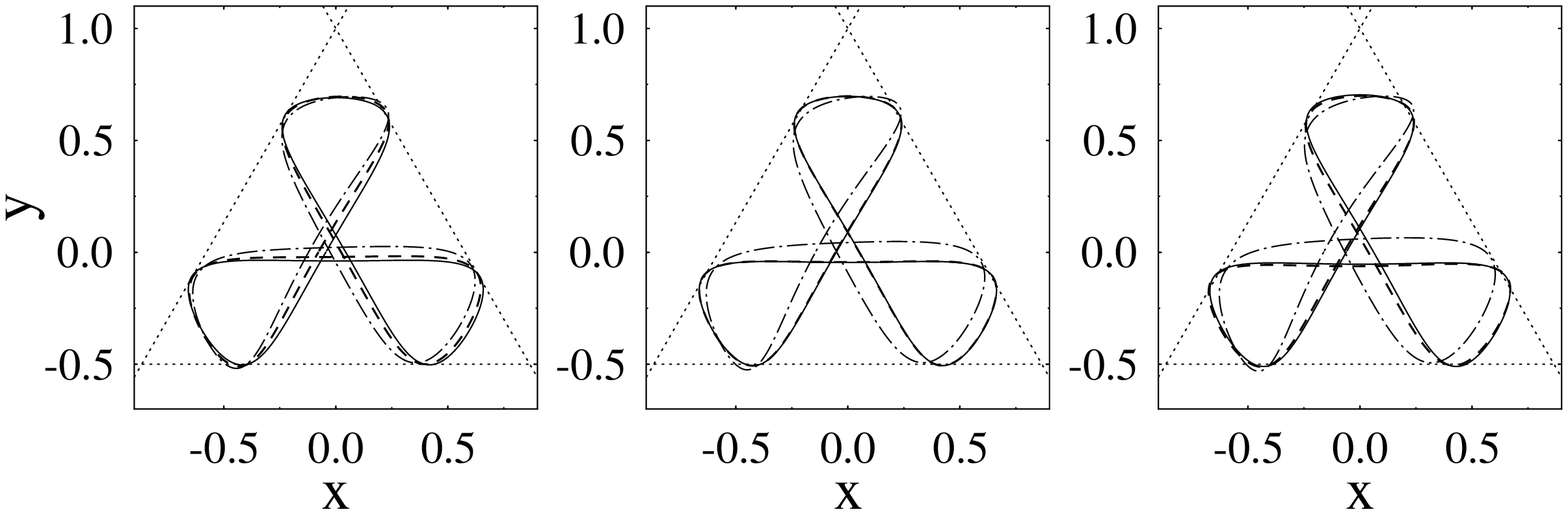}{3.9}{14.5}{
Shapes of orbits D$_7$/D$_9$ (solid), E$_8$ (dashed), and G$_7$ 
(dash-dotted lines) in the $(x,y)$ plane, shown at the three 
energies $e=1.170$ (left panel), $e=1.179$ (center panel), and  
$e=1.190$ (right panel). The dotted lines are the equipotential 
lines at $e=1$ intersecting at the three saddle points.
}

\vspace*{-0.2cm}
Since the Gutzwiller trace formula \eq{dggamma} diverges at the 
bifurcation points, so-called uniform approximations \cite{ozoha,ssuni} 
must be used. For the codimension-two scenario described above, a suitable
uniform approximation has been developed by Schomerus \cite{scho2}; we have
adapted it to the present scenario (see \cite{kaid} for the technical 
details). Expanding the normal form used in the trace integral near the two 
bifurcations \cite{scho2}, the traces of the stability matrices of the 
participating orbits are given by
\be
{\rm tr}\,{\widetilde M}_{\,\rm E,G} 
                 = 2 \mp 6a\sqrt{-\epsilon_1\!/3a} 
                   - \frac{8b}{3a}\,\epsilon_1 
                   + {\cal O}(|\epsilon_1|^{3/2})\,,
\label{loctr1}
\ee
with $\epsilon_1=c_1(e-e_1)$ near the first bifurcation, and by
\bea
{\rm tr}\,{\widetilde M}_{\,\rm D} & = & 2 - \epsilon_2^2\,,\nonumber\\
{\rm tr}\,{\widetilde M}_{\,\rm E} & = & 2 + 3\epsilon_2^2
                       + {\cal O}\left(\epsilon_2^4\right),
\label{loctr2}
\eea
with $\epsilon_2=c_2(e-e_2)$ near the second bifurcation. The behaviour
of tr${\widetilde M}_\xi(e)$ given by the above formulae  near the two 
bifurcations, i.e.\ for $\epsilon_1\ll 1$ and $\epsilon_2\ll 1$, can 
clearly be recognized in \fig{trmbif}, where the solid lines represent 
the real parts of tr${\widetilde M}_\xi(e)$. The short-dashed lines shown 
for $e\leq e_1$ represent the real part (upper left) and imaginary parts 
(lower left) of tr${\widetilde M}_{E',G'}(e)$ corresponding to the complex 
continuations of the periodic orbits E$_8$ and G$_7$ taking part in the 
tangent bifurcation, i.e., the so-called ``ghost orbits'', denoted here by E' 
and G'. They have to be included in order to obtain a continuous description
of the semiclassical density of states throughout the whole bifurcation
region. The parameters $a<0\,$, $b<0\,$, $c_1>0\,$, and $c_2>0$ appearing
in Eqs.\ \eq{loctr1} and \eq{loctr2} come from the normal form of the trace 
integral, given in \cite{scho2}, and depend on the system. They need, 
however, not be determined explicitly but can be expressed in terms of the 
numerically calculated quantities $S_\xi$ and tr${\widetilde M}_\xi(e)$ of 
the periodic orbits \cite{scho2,kaid}.

\Figurebb{cont}{0}{-10}{355}{106}{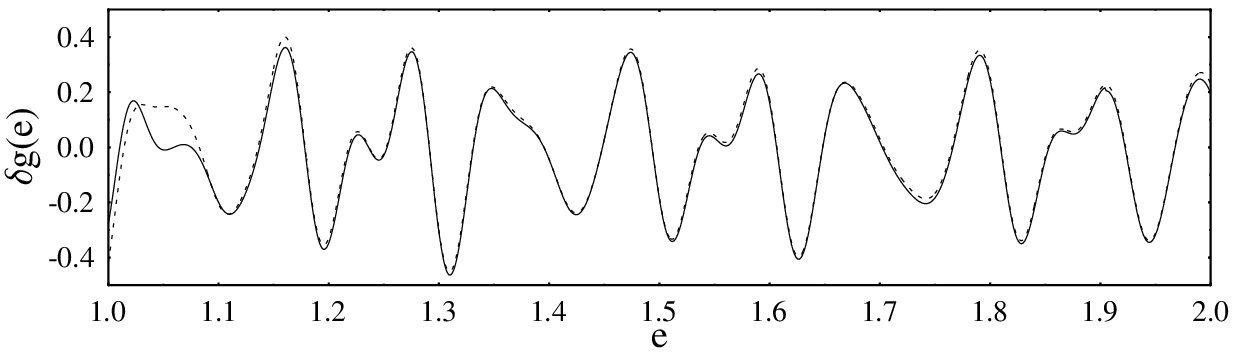}{4.0}{14.5}{
Oscillating part of the coarse-grained density of states of the HH system 
with $\alpha=0.1$. The quantum result is shown by the solid, the 
semiclassical result by the dashed line. Gaussian smoothing range: 
$\gamma=0.5$. The primitive orbits C$_3$, B$_4$, R$_5$, L$_6$, and the first 
three repetitions of $\tau_2$ are included in the trace formula \eq{dggamma}.
}

In \fig{cont} we show a comparison between the quantum-mechanical and the 
semiclassical result for the oscillating part $\delta g_\gamma(E)$ of the
coarse-grained density of states in the energy region $1\leq e \leq 2$. 
Hereby an energy averaging width $\gamma=0.5$ has been used. In the
semiclassical calculation, only the primitive orbits C$_3$, B$_4$, R$_5$, 
L$_6$, and the first three repetitions of $\tau_2$ had to be included for
this energy resolution; no bifurcation occurs for these orbits so that the 
original trace formula for isolated orbits \eq{dggamma} could be used.
The agreement is seen to be very good except in the region just above 
$e=1$. The small discrepancies seen there are due to the plateau 
uncertainties in the Strutinsky averaging of the smooth density of states,
which have been mentioned at the end of Sec.\ \ref{secstru}.

\Figurebb{cont3}{0}{-10}{355}{106}{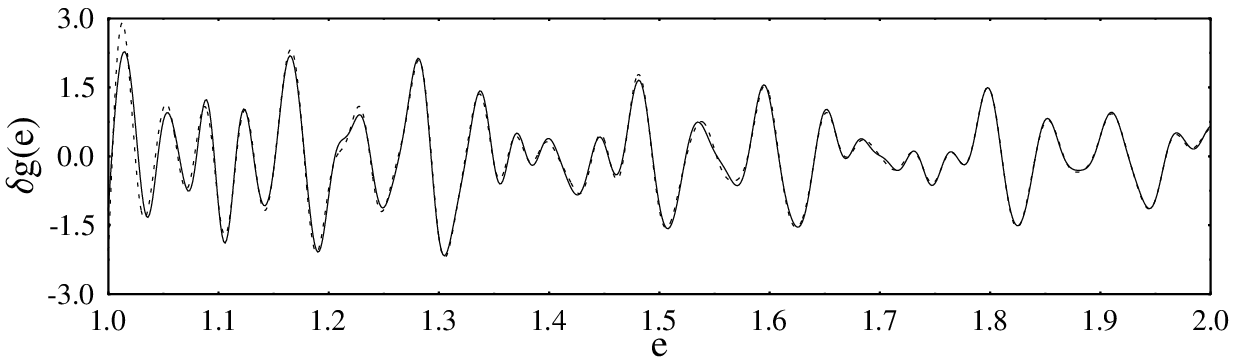}{4.0}{14.5}{
Same as \fig{cont}, but with Gaussian smoothing range $\gamma=0.25$.
Here also R$_7$ and L$_8$, the second repetitions of B$_4$ and C$_3$, up 
to five repetitions of the orbit $\tau_2$, and the period-two orbits
D$_7$/D$_9$, E$_8$ and G$_7$ are included; the latter three in the uniform 
approximation discussed in the text.}

In \fig{cont3} the same comparison is made, but this time with a finer
energy resolution given by $\gamma=0.25$. To obtain convergence of the
trace formula, also the period-two orbits D$_7$/D$_9$, E$_8$ and G$_7$ had
to be included (in the uniform approximation mentioned above), besides the
orbits R$_7$ and L$_8$, the second repetitions of B$_4$ and C$_3$, and up 
to five repetitions of the orbit $\tau_2$. Again the agreement between
quantum mechanics and semiclassics is nearly perfect, in spite of the 
rather complex gross-shell structure in $\delta g(E)$. The uncertainties 
in $\widetilde{g}(E)$ here have less relative weight than for $\gamma=0.5$, 
due to the larger overall amplitude of the oscillations in $\delta g(E)$. 

Finally, in \fig{cont3us} we show the results obtained with exactly the
same parameters as in \fig{cont3}, but this time omitting the stable
orbit D$_7$/D$_9$ and its companions E$_8$ and G$_7$ involved in the
bifurcation. The difference to \fig{cont3} is not large, but it shows
that the inclusion of a stable bifurcating orbit in this mixed system
does improve the semiclassical description.

\Figurebb{cont3us}{15}{-10}{355}{106}{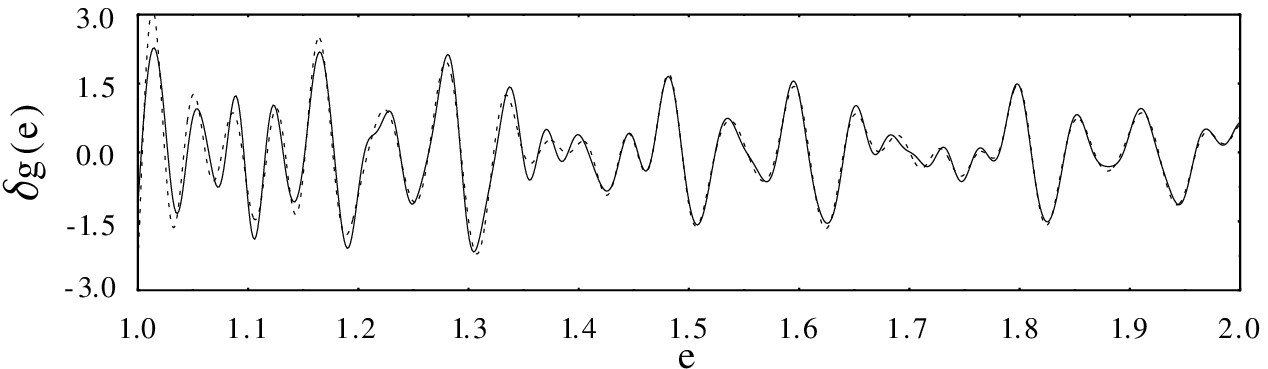}{4.0}{14.5}{
Same as \fig{cont3}, but omitting the orbits D, E, and G participating in the 
bifurcations.}

The process of diminishing $\gamma$ and including longer periodic orbits
could, in principle, be continued -- but at the cost of having to deal
with more and more bifurcations, including scenarios of codimension
higher than two. This just expresses the practical impossibility of
obtaining full semiclassical quantisation in a mixed system, which we
already discussed in the introduction. Here we have demonstrated, however, 
that the slightly {\it coarse-grained} density of states can be well 
described semiclassically using a relatively small number of periodic
orbits, one of which (D$_7$/D$_9$) is stable in a sizeable part of the
energy region explored.

\section{Summary and conclusions}

We have presented the first semiclassical calculation of the density of 
states in the H\'enon-Heiles (HH) potential in the high-energy continuum 
region up to twice the barrier energy ($e=2$). Above the barrier ($e>1$), 
the density of states is dominated by resonances with appreciable imaginary 
parts. We have calculated the quantum-mechanical resonance spectrum using 
the method of complex rotation. Even though the HH system is classically 
almost chaotic for $e>1$, it still possesses small stable islands and 
therefore exhibits the general problems with the semiclassical quantization
that are characteristic of mixed systems. We have shown, however, that the 
slightly coarse-grained 
density of states with a finite energy resolution $\gamma$ can very well be 
approximated semiclassically in terms of relatively few stable and unstable
periodic orbits. For an energy resolution of $\gamma=0.25$ (see \fig{cont3}), 
we had to use only 11 orbits; three of them as a cluster taking part in a 
codimension-two bifurcation, and the rest as isolated orbits. The agreement 
between the semiclassical and the quantum-mechanical results for the 
oscillating part $\delta g_\gamma(E)$ of the density of states is excellent, 
except near the barrier energy ($e\simg 1$) where it is difficult to extract 
a unique average part $\widetilde{g}(E)$ of the density of states. 
While the (extended) Thomas-Fermi model fails completely, due to the 
unboundedness of the system for $e>1$, we have successfully implemented the 
numerical Strutinsky-averaging technique to obtain the average part. Some 
minor uncertainties in its plateau value remaining near $e\simg 1$ play a 
decreasing relative role with increasing energy resolution, i.e., with 
decreasing coarse-graining width $\gamma$.
 
We conclude that the semiclassical trace formula, complemented by the uniform 
treatment of bifurcating periodic orbits, is a very economic tool for 
prediciting quantum oscillations also in the continuum region of a 
mixed-dynamical Hamiltonian system which is dominated by scattering resonances.

\bs

We are grateful to Gregor Tanner for a careful reading of the manuscript
and for clarifying comments.
This work was supported by the Deutsche Forschungsgemeinschaft (DFG) via 
the graduate college 638 ``Nonlinearity and Nonequilibrium in Condensed 
Matter''. One of the authors (P.W.) gratefully acknowledges travel support 
from the Office of Academic Affairs and the Foundation of the University 
of Nevada at Reno.

\section{Appendix: Strutinsky averaging of the resonance spectrum}

The Strutinsky-averaged density of resonances is, according to Eqs.\ 
\eq{dos}, \eq{fgam} and \eq{gav}, defined by
\begin{eqnarray}
\widetilde{g}(E) = \frac{1}{\,\tilde{\!\gamma}\sqrt{\pi}} 
                       \int_{-\infty}^{\infty}
                       e^{-\left[(E-E')/\tilde{\!\gamma}\right]^2}
                       L^{\frac{1}{2}}_s \!\left[\left(\frac{E-E'}
                       {\,\tilde{\!\gamma}}\right)^{\!\!2}\right] \frac{1}{\pi}
                       \sum_m \frac{\Gamma_{\!m}/2}{\left(E'-E_m\right)^2
                       +(\Gamma_{\!m}/2)^2}\,\d E'. 
\label{strbw}
\end{eqnarray}
Without the Laguerre
polynomial $L^{\frac{1}{2}}_s$, the integral in \eq{strbw} is identical
to that in \eq{ggam} and can be expressed in terms of the quantities
given in \eq{erf} with the substitution $\gamma\rightarrow\,\tilde{\!\gamma}$.
The Laguerre polynomials with $s>0$ create even powers of 
$(E-E')/\,\tilde{\!\gamma}$ under the integral. These can be included using 
the formula
\begin{eqnarray}
\int_{-\infty}^{\infty}\left(\frac{E-E'}{\,\tilde{\!\gamma}}\right)^{\!k}
    \frac{\exp[-(E-E')^2\!/\,\tilde{\!\gamma}^2]}
    {(\Gamma_{\!m}/2)^2+\left(E_m-E'\right)^2}\,\d E'
  = \frac{2 \pi (-1)^k}{\Gamma_{\!m}} {\rm Re}
    \left\{\frac{d^{k/2}}{d \lambda^{k/2}} \; w\!\left[\frac{\sqrt{\lambda}}
    {\,\tilde{\!\gamma}}\left(E_m+i\frac{\Gamma_{\!m}}{2}-E\right)\right]
    \right\}_{^{\!\lambda=1}}\nonumber
\end{eqnarray}
with $k=0,2,\dots,2s$. The final expressions for $\widetilde{g}(E)$ obtained 
in this way for $s\leq 5$ are given in \tab{avdos}.

\ms

\begin{table}[H]
\begin{center}
\renewcommand{\arraystretch}{1.5}
\begin{tabular}{|c|l|}
{\large ${\displaystyle \;s\;}$} & 
{\large ${\displaystyle
\,\tilde{\!\gamma}\sqrt{\pi}\,\widetilde{g}(E)}$} \\[1mm] \hline \hline
$0$ & $\sum_m \Re e \; w\left(z_m\right)$ \\[1mm] \hline
$1$ & $\sum_m \Re e \left(L^{\frac{1}{2}}_1\!\left(z^2_m\right) 
w\left(z_m\right)+i\,z_m/\!\sqrt{\pi}\right)$ 
\\[1mm] \hline 
$2$ & $\sum_m \Re e \left(L^{\frac{1}{2}}_2\!\left(z^2_m\right) 
w\left(z_m\right)-i\, z_m \left(2z^2_m-9\right)\!/4 \sqrt{\pi}
\right)$ \\[1mm] \hline 
$3$ & $\sum_m \Re e \left(L^{\frac{1}{2}}_3\!\left(z^2_m\right) 
w\left(z_m\right)+i\, z_m 
\left(4z^4_m-40z^2_m+87\right)\!/24\sqrt{\pi}\right)$ \\[1mm] \hline 
$4$ & $\sum_m \Re e \left(L^{\frac{1}{2}}_4\!\left(z^2_m\right) 
w\left(z_m\right)-i\,z_m 
\left(8z^6_m-140z^4_m+690z^2_m-975\right)\!/192 \sqrt{\pi}\right)$ \\[1mm] \hline 
$5$ & $\sum_m \Re e \left(L^{\frac{1}{2}}_5\!\left(z^2_m\right) 
w\left(z_m\right)+i\,z_m 
\left(16z^8_m-432z^6_m+3752z^4_m-12180z^2_m+12645\right)\!/1920\sqrt{\pi}\right)
$\hspace*{4.5cm}\\[1mm]
\end{tabular}
\parbox{14.5cm}{
\vspace{0.5cm}
\caption[table]{\renewcommand{\baselinestretch}{0.8} \small
                                           \hspace{-0.3truecm}
Strutinsky-averaged density of states \eq{strbw} for orders $0\leq s\leq 5$ of
the correction polynomial. The definitions of $z_m$ and $w(z)$ are given in
Eq.\ \eq{erf} with $\gamma$ replaced by $\,\tilde{\!\gamma}$.
}
\label{avdos}}
\end{center}
\end{table}

\end{document}